\documentclass[aps,amssymb,prl,twocolumn,superscriptaddress]{revtex4}
\usepackage{graphicx}

\begin{document}

\title{Magnon-Hole Scattering and Charge Order in $\rm\bf Sr_{14-x}Ca_xCu_{24}O_{41}$}

\author{C. Hess}
\email[]{christian.hess@physics.unige.ch}

\affiliation{DPMC, Universit\'{e} de Gen\`{e}ve, 1211 Gen\`{e}ve,  Switzerland}
\affiliation{2. Physikalisches Institut, RWTH-Aachen, 52056 Aachen, Germany}
\author{H. ElHaes}
\affiliation{2. Physikalisches Institut, RWTH-Aachen, 52056 Aachen, Germany}
\author{B. B\"{u}chner}
\affiliation{2. Physikalisches Institut, RWTH-Aachen, 52056 Aachen, Germany}
\affiliation{Leibniz-Institute for Solide State and Materials Research, IFW-Dresden, 01171 Dresden, Germany}

\author{U. Ammerahl}
\affiliation{Laboratoire de Physico-Chimie de l'\'{E}tat Solide, Universit\'{e} Paris-Sud, 91405 Orsay, France}

\author{M. H\"{u}cker}
\affiliation{Laboratoire de Physico-Chimie de l'\'{E}tat Solide, Universit\'{e} Paris-Sud, 91405 Orsay, France}
\affiliation{Physics Department, Brookhaven National Laboratory, Upton, New York 11973 }

\author{A. Revcolevschi}
\affiliation{Laboratoire de Physico-Chimie de l'\'{E}tat Solide, Universit\'{e} Paris-Sud, 91405 Orsay, France}

\date{\today}

\begin{abstract}
The magnon thermal conductivity $\kappa_{\mathrm{mag}}$ of the hole doped spin ladders in $\rm
Sr_{14-x}Ca_xCu_{24}O_{41}$ has been investigated at low doping levels $x$. The analysis of $\kappa_{\mathrm{mag}}$ reveals a strong doping and temperature dependence of the magnon mean free path $l_{\mathrm{mag}}$ which is a local probe for the interaction of magnons with the doped holes in the ladders. In particular, this novel approach to studying charge degrees of freedom via spin excitations shows that charge ordering of the holes in the ladders leads to a freezing out of magnon-hole scattering processes.
\end{abstract}


\maketitle

The interplay between charge and spin degrees of freedom is a crucial aspect of the physics of
transition metal oxides. For example, the pairing mechanism in high-temperature superconductors is most likely related to magnetic excitations. Moreover, the competition between charge mobility and magnetic interactions appears to be the source of charge ordering phenomena \cite{Kivelson98,Moreo99,Zaanen99}, such as stripes in cuprates and nickelates or phase separation
in manganites. Experimental studies yield in particular clear-cut evidence that static stripe order and superconductivity are competing ground states in two dimensional cuprates \cite{Zaanen99,Klauss00}. A similar competition has been predicted from theoretical analysis of hole doped S=1/2 spin ladders as a one-dimensional model system \cite{Dagotto92,Dagotto96}. Such hole-doped spin ladders are realized in the compound $\rm Sr_{14-x}Ca_xCu_{24}O_{41}$, where via Ca-doping the hole-doping level in the ladders can be controlled since holes are redistributed between the ladders and the also-present spin chains \cite{Nucker00,Osafune97}.
At first sight, the available experimental data on this material seem to support the predicted scenario of competing charge ordering and superconducting ground states: charge order, which has been reported in the chemically undoped compound $\rm Sr_{14}Cu_{24}O_{41}$ \cite{Takigawa98,Regnault99,Kataev01,Blumberg02,Gorshunov02}, gradually destabilizes upon Ca-doping \cite{Matsuda99,Kataev01,Vuletic03a} and superconductivity eventually occurs at high doping levels if high external pressure is applied \cite{Uehara96}.
However, it is difficult to experimentally prove the existence of charge order in the ladders, since $\rm Sr_{14-x}Ca_xCu_{24}O_{41}$ also contains hole doped spin chains, whose charge ordered ground state is well established \cite{Takigawa98,Regnault99,Kataev01}. Moreover, no direct information currently exists concerning the interplay between charge dynamics and magnetic excitations in the ladders.

In general, transport experiments are an excellent tool for investigating the interplay between different degrees of freedom since they probe the scattering and dissipation of excitations. Electron-phonon, electron-electron and electron-magnon scattering, for example, can be studied via electrical resistivity $\rho$. In the case of charge ordering, however, the drastic change of charge mobility rather than scattering dominates the anomalies of electrical transport. We therefore use a novel approach and study electronic degrees of freedom via the electron-magnon interaction by measuring the transport and scattering of magnetic excitations. As has been demonstrated in Ref.~\cite{Hess01,Sologubenko00,Hess03}, the thermal conductivity $\kappa$ is a valuable tool for this purpose, since magnetic contributions to this quantity yield the mean free path of magnetic excitations. 
In $\rm Sr_{14-x}Ca_xCu_{24}O_{41}$, magnon heat transport along the ladders generates a strong anisotropy in the $\kappa$ tensor. While conventional phonon heat conduction is observed perpendicular to the ladders,  $\kappa$ is much larger and often exhibits a high-temperature peak due to magnon contributions \cite{Hess01,Sologubenko00,Hess02}.

In this letter we utilize this magnon heat transport to selectively investigate the interaction between magnons and holes in the ladders by means of magnon hole scattering. We show that the temperature dependence of the magnon mean free path $l_{\mathrm{mag}}$ in the ladders is unambiguously correlated with the mobility of holes. Charge ordering in the ladders is accompanied by a drastic enhancement of $l_{\mathrm{mag}}$: the scattering probability, which is close to unity for mobile holes, vanishes in the charge ordered state.

We have measured $\kappa$ and $\rho$ of $\rm Sr_{14-x}Ca_xCu_{24}O_{41}$
($x=0$, 2, 3, 4, 5) single crystals with standard four probe techniques \cite{Hess99,Hess01}. Details of the sample preparation were published in Ref.~\cite{Ammerahl98}.


\begin{figure}
\includegraphics [width=.6\columnwidth,clip] {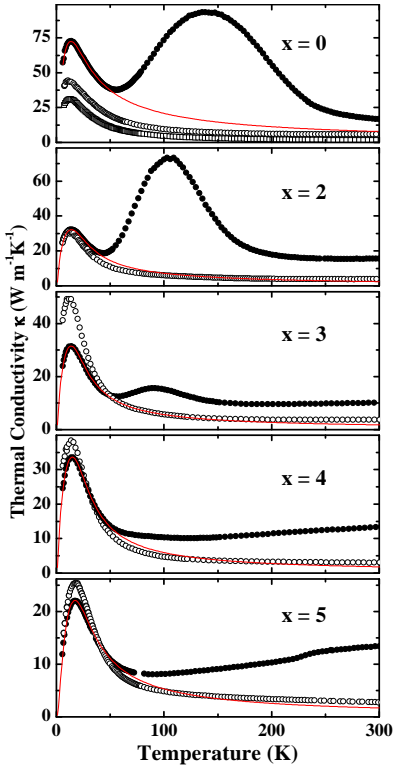}
\caption{\label{fig1}$\kappa_c$ ($\bullet$) and $\kappa_a$ ($\circ$) of $\rm
Sr_{14-x}Ca_xCu_{24}O_{41}$ ($x=0$, 2, 3, 4, 5) as a function of temperature. Solid lines represent
the estimated phononic background of $\kappa_c$. For $x=0$ also $\kappa$ along the $b$-axis is shown ($\square$).}
\end{figure}

In Fig.~\ref{fig1} we present the $T$-dependence of the thermal conductivity parallel ($\kappa_c$) and
perpendicular ($\kappa_a$, $\kappa_b$) to the ladders of $\rm Sr_{14-x}Ca_xCu_{24}O_{41}$ ($x=0$, 2, 3, 4, 5). The signature of the one-dimensional magnetic heat transport is most evident for $x=0$. Here, a typical  phonon thermal conductivity is found for $\kappa_a$ and $\kappa_b$. In addition to such a phononic background in $\kappa_c$, a huge peak due to magnons is present at higher $T$. Similar magnon peaks are also observed for the Ca-doped materials with $x\leq3$. It is, however, evident that with increasing $x$ the peak shifts gradually towards lower $T$ and its maximum strongly decreases. It is eventually absent completely for $x=4$ and $x=5$. Nevertheless, there is a pronounced anisotropy between $\kappa_a$ and $\kappa_c$ also at this higher doping level: for $T\gtrsim20$~K $\kappa_a$ monotonically decreases whereas $\kappa_c$ exhibits a minimum around 100~K and increases with further rising $T$, exceeding the phononic $\kappa_a$ by up to a factor of five at room temperature. We thus conclude that magnon heat
conduction still contributes significantly to $\kappa_c$ also in these cases \cite{Hess02}.

\begin{figure}
\includegraphics [width=\columnwidth,clip] {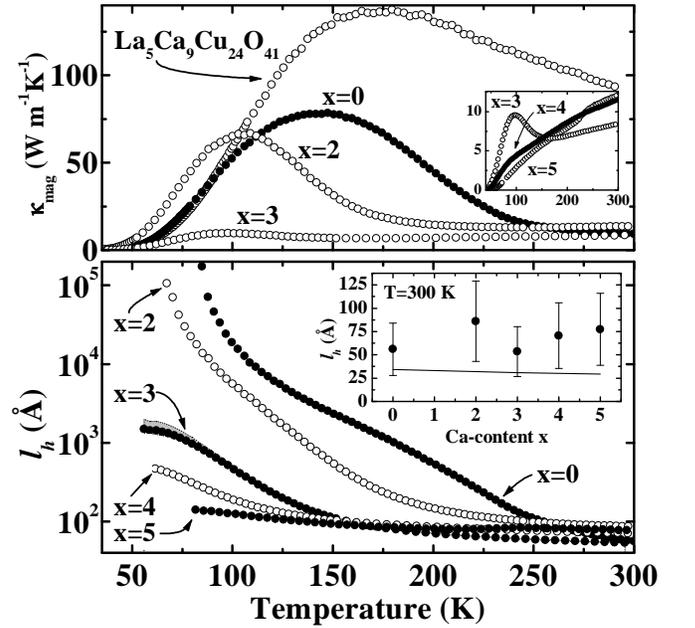}
\caption{\label{fig2}Top: $\kappa_{\mathrm{mag}}(T)$ of $\rm Sr_{14-x}Ca_xCu_{24}O_{41}$ ($x=0$, 2, 3, 4,
5) and of $\rm La_5Ca_9Cu_{24}O_{41}$. Inset: enlarged representation for $x=3$, 4, 5.
Bottom: $l_h(T)$ for $x=0$, 2, 3, 4, 5. The effect of possible errors in $l_0$ (cf. text) is shown as an example for $x=3$ by the shaded area. Inset: doping dependence of $l_h$ at 300~K. The error
bars arise due to an estimated error of 50\% for the phonon background. The solid line
represents the mean hole distance as calculated from Ref.~\cite{Nucker00}.}
\end{figure}

Due to the large spin gap \cite{Eccleston98,Katano99} the magnon thermal conductivity $\kappa_{\mathrm{mag}}$ is negligible below
$T\lesssim40$~K \cite{Hess01}. This can be used to separate the phonon and magnon contributions.
We fitted $\kappa_c$ for $T\lesssim40$~K with a usual Debye model \cite{Callaway60} and extrapolated
this fit up to $T=300$~K in order to obtain the phonon thermal conductivity $\kappa_{\mathrm{ph}}$ (solid lines in Fig.~\ref{fig1}). Subtraction of $\kappa_{\mathrm{ph}}$ from $\kappa_c$ yields $\kappa_{\mathrm{mag}}$ which is shown in the top panel of Fig.~\ref{fig2}.

Prior to discussing the effect of Ca-doping on $\kappa_{\mathrm{mag}}$ we briefly review the
differences between $\kappa_{\mathrm{mag}}$ of $\rm Sr_{14}Cu_{24}O_{41}$ and
$\kappa_{\mathrm{mag}}$ of $\rm La_5Ca_9Cu_{24}O_{41}$, which is also shown in the top panel of
Fig.~\ref{fig2} \cite{Hess01}. It is known from spectroscopic experiments that the La-based compound contains undoped ladders, in contrast to $\rm Sr_{14}Cu_{24}O_{41}$ where the hole-content in the ladders is finite \cite{Osafune97,Nucker00}. The effect of hole-doping on $\kappa_{\mathrm{mag}}$ can therefore immediately
be observed. For $T \lesssim100$~K $\kappa_{\mathrm{mag}}$ increases with $T$ almost identically for both compounds. Pronounced differences occur only at higher $T$: $\kappa_{\mathrm{mag}}$ of $\rm La_5Ca_9Cu_{24}O_{41}$ exhibits a large peak ($\sim140~\rm Wm^{-1}K^{-1}$ at $\sim180$~K) and stays very large even at room temperature ($\sim100~\rm Wm^{-1}K^{-1}$). In contrast, the peak is much smaller in the case of $\rm Sr_{14}Cu_{24}O_{41}$ ($\sim75~\rm Wm^{-1}K^{-1}$ at $\sim150$~K). Here $\kappa_{\mathrm{mag}}$ decreases much more strongly at high $T$ and saturates at $\kappa_{\mathrm{mag}}\approx10~\rm Wm^{-1}K^{-1}$ for $T\gtrsim240$~K.

The strong suppression of $\kappa_{\mathrm{mag}}$ in $\rm Sr_{14}Cu_{24}O_{41}$ at high temperatures must be related to scattering of the magnons on holes, since structural defects due to different ions on the Sr-site are known to not influence $\kappa_{\mathrm{mag}}$ and therefore the hole doping is the only difference with respect to the undoped ladders. Since both $\kappa_{\mathrm{mag}}$ curves are almost identical below $T_0\approx100$~K, this scattering mechanism obviously becomes completely unimportant below $T_0$ and unfurls its full strength above a characteristic temperature $T^*\approx 240$~K. The comparison with $\kappa_{\mathrm{mag}}$
of the Ca-doped samples (see Fig.~\ref{fig2}, top panel) reveals that $T_0$ and $T^*$ are gradually shifted
towards lower $T$; i.e., the temperature region where $\kappa_{\mathrm{mag}}$ is suppressed extends
and magnon-hole-scattering also becomes important at low $T$.
Apparently, at $x=4$, 5 this region becomes so wide that even the peak at low $T$ is suppressed.

In order to elucidate the origin of this interesting observation we have analyzed
$\kappa_{\mathrm{mag}}$ of $\rm Sr_{14-x}Ca_xCu_{24}O_{41}$ and calculated $l_{\mathrm{mag}}$ as a
function of $T$. In Ref.~\cite{Hess01} $\kappa_{\mathrm{mag}}$ of $\rm Sr_{14}Cu_{24}O_{41}$ and
$\rm La_5Ca_9Cu_{24}O_{41}$ was analyzed using a kinetic model. A different analysis of the data based on a microscopic model leads to qualitatively similar results, though the values of $l_{\mathrm{mag}}$ are smaller \cite{Alvarez02}. We nevertheless use the simple kinetic approach in the following analysis, since more
recent calculations \cite{foot1} differ from the results in \cite{Alvarez02}. Moreover, preliminary studies of impurity
effects strongly support the larger $l_{\mathrm{mag}}$ reported in \cite{Hess01}.

Firstly, $\kappa_{\mathrm{mag}}$ is fitted at low $T$ with
\begin{equation}\label{fitmag}
\kappa_{\mathrm{mag}}=\frac{3 N l_{\mathrm{mag}}}{\pi\hbar
k_BT^2}\int_{\Delta_{\mathrm{ladder}}}^{\epsilon_{\mathrm{max}}}
\frac{\exp(\epsilon/k_BT)}{(\exp(\epsilon/k_BT)+3)^2}\epsilon^2d\epsilon
\end{equation}
where $N$ is the number of ladders per unit area and $\epsilon_{\mathrm{max}}\approx200$~meV is the band
maximum of the spin excitations \cite{Eccleston98}. This yields the spin gap $\Delta_{\mathrm{ladder}}$ and $l_0$, i.e., the low-$T$ value of $l_{\mathrm{mag}}$ which is assumed to be constant in the fitting range \cite{Hess01}. $l_{\mathrm{mag}}=l_0$ arises when scattering on quasiparticles freezes out and magnon-defect
scattering dominates. The second step of the analysis comprises the calculation of $l_{\mathrm{mag}}(T)$
by comparing experimental and theoretical data of $\kappa_{\mathrm{mag}}$ using Eq.~\ref{fitmag} with a known value for $\Delta_{\mathrm{ladder}}$. Due to the strong suppression of $\kappa_{\mathrm{mag}}$ in the higher Ca-levels, such a determination of $\Delta_{\mathrm{ladder}}$ and $l_0$ is only reasonable for $x\leq 2$.
Therefore, we use $\Delta_{\mathrm{ladder}}/k_B=377$~K as known from neutron scattering \cite{Eccleston98,Katano99} for the calculation of $l_{\mathrm{mag}}(T)$ for $x\geq3$. 

In order to separate scattering effects due to holes from the total $l_{\mathrm{mag}}$ we
apply Matthiesen's rule $ 1/{l_{\mathrm{mag}}}={1}/{l_0}+1/{l_h}$, where $l_h$ denotes the
hole-scattering part of $l_{\mathrm{mag}}$ and is a measure for the importance of magnon-hole
scattering \cite{foot2a}. $l_h$ is related to the mean distance of holes $d_h$ via the effective scattering probability $\gamma_h$ by $l_h=d_h/\gamma_h$. While $l_0$ is known from our fits of the low-$T$ increase of $\kappa_{\mathrm{mag}}$ for $x\leq2$, $l_0=3000\pm1000$~{\AA} has been assumed \cite{foot2}.

$l_h(T)$ of $\rm Sr_{14-x}Ca_xCu_{24}O_{41}$ is depicted in the bottom panel of Fig.~\ref{fig2}.
Obviously, $l_h$ systematically decreases with increasing Ca-content which is strong evidence
for the growing importance of magnon-hole scattering upon Ca-doping, i.e., with increasing hole content in the ladders. All curves also decrease with
rising $T$ and saturate around 50-80~{\AA}, which is of the same order of magnitude as the mean
distance of holes $d_h\approx30$~{\AA} \cite{Nucker00} (cf. inset of Fig.~\ref{fig2}). At room temperature magnons appear to be strongly scattered on holes with a scattering probability $\gamma_h=d_h/l_h=0.5-1$, which is also consistent with the pronounced suppression of $\kappa_{\mathrm{mag}}$ above $T^*$. Note that in this high-$T$ range the uncertainty in $\kappa_{\mathrm{ph}}$ leads to a relative error in $l_h$ of about 50\% which explains the apparent non systematic high-$T$ behavior of $l_h$.

\begin{figure}
\includegraphics [width=\columnwidth,clip] {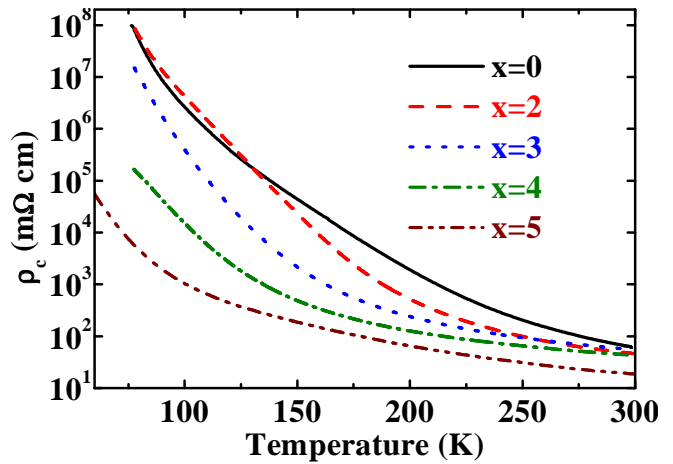}
\caption{\label{fig3} Temperature dependence of the electrical resistivity along the ladders $\rho_c$ of $\rm Sr_{14-x}Ca_xCu_{24}O_{41}$ ($x=0$, 2, 3, 4, 5).}
\end{figure}

\begin{figure}
\includegraphics [width=\columnwidth,clip] {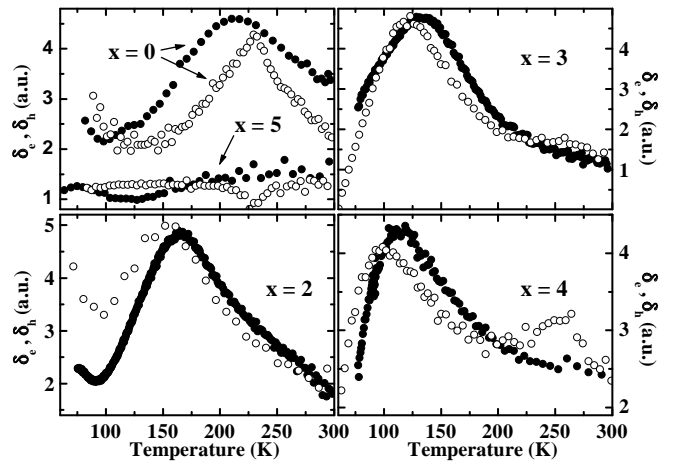}
\caption{\label{fig4} Temperature dependence of $\delta_h$ ($\circ$) and $\delta_e$ ($\bullet$) in $\rm Sr_{14-x}Ca_xCu_{24}O_{41}$ ($x=0$, 2, 3, 4, 5).}
\end{figure}

It is possible to read off the characteristic temperature $T^*$ from these curves for $x\leq3$,
since it separates an almost constant high-$T$ behavior from a steep $T$-dependence at low $T$. This $T$-dependence is clearly correlated with the electrical resistivity $\rho$ as is evident from our measurements along the $c$-axis ($\rho_c$) depicted in Fig.~\ref{fig3}. For $x \leq 2$ the highly similar $T$-dependence of $\rho_c$ and $l_h$ is apparent \cite{foot3}.
In order to show this correlation of $l_h$ and $\rho$ also for higher doping levels, we compare in Fig.~\ref{fig4} the logarithmic derivatives of $l_h$ and $\rho_c$, i.e., $\delta_h=\frac{d}{d(1/T)} \ln l_h$ (open circles) and $\delta_e=\frac{d}{d(1/T)}\ln\rho_c$ (full circles) \cite{foot4}. As is obvious from the figure, $\delta_h$ and $\delta_e$ exhibit a very similar $T$-dependence for all $x$. This proves unambiguously that the magnon heat transport and the electric transport are closely linked to each other and that charge degrees of freedom are by far the strongest scatterers of spin excitations. Since the peaks in $\delta_e$ are signatures of the charge ordering in this material \cite{Carter96}, the freezing-out of magnon-hole scattering has to be attributed to the formation of this charge ordered state.

It is however very surprising that this scattering channel vanishes completely in the charge ordered state. In the framework of our model there are two possible scenarios to interprete this intriguing finding. Since $\kappa_{\mathrm{mag}}$ selectively probes the magnetic excitations in the ladders, either the relevant hole concentration in the ladders decreases drastically below $T^*$, or the scattering probability $\gamma_h$ vanishes upon charge ordering {\em in the ladders}.

The first scenario seems to be supported by a recent $^{17}\rm O$-NMR study \cite{Thurber03}. A strong change of the electrical field gradient is observed below $T^*$ which the authors interpret in terms of a complete transfer of holes from the ladders to the chains below $T^*$. This interpretation, however, is not supported by preliminary x-ray absorption spectroscopy (XAS) data on these compounds \cite{Nucker03}, which clearly show that the spectra of the charge ordered state differ drastically from the spectra of undoped ladders. The XAS data do indeed exhibit some clearcut changes at $T^*$; however, the change of the hole distribution between  ladders and chains as signalled by these data is only subtle, if present at all. 

The second scenario, i.e., the reduction of the scattering probability at constant hole content, is reasonable if the charge order is connected with a periodic modulation of the spin density. Though possible, it seems unlikely that the periodic arrangement of the holes is the only reason for the strong suppression of scattering. In order to be compatible  with the large $\kappa_{\mathrm{mag}}$ and $l_{\mathrm{mag}}$ at low $T$ the charge order would have to be perfect on a unrealistically large lengthscale with a correlation length $\xi\gtrsim l_0$. 

Further studies are necessary to clarify the origin of the drastic change of $l_{\mathrm{mag}}$ at $T^*$. For example, one might speculate that of hole pair formation and/or a change of the orbital character of the ladder's hole states, as could be signalled by the NMR and XAS data, play a relevant role in the magnetic heat transport.
Besides these uncertainties, however, there are also clear-cut conlusions to be made from our data. Firstly, charge ordering is indeed present in the ladders of $\rm Sr_{14-x}Ca_xCu_{24}O_{41}$. This charge ordering is strongly linked to magnetic degrees of freedom. Finally, measuring the magnon heat transport is a valuable tool to study the interaction of charge and spin degrees of freedom.

In conclusion, we have shown that the magnon heat conductivity $\kappa_{\mathrm{mag}}$ of $\rm Sr_{14-x}Ca_xCu_{24}O_{41}$ is strongly affected by magnon-hole scattering. Drastic temperature and doping dependent changes have been found to be clearly correlated with anomalies of electronic transport. Our data suggest charge ordering in the ladders, which has a strong impact on magnetic degrees of freedom. 

This work has been supported by the DFG through SP1073. C.H. acknowledges support by the DFG through HE3439/3-1. M.H. acknowledges support by the U.S. Department of Energy, under Contract No. DE-AC02-98CH10886. We further thank W. Brenig and F. Heidrich-Meisner for stimulating discussions and A.P. Petrovic for proofreading the manuscript.


\begin{thebibliography}{31}
\bibitem{foot2a}We should note, that no proof exists for temperature-independent magnon-defect scattering ($l_0$) and the applicability of Mathiesen's rule, yet this is consistent with previous results \cite{Hess01}.
\bibitem{foot2}{We emphasize that $l_{\mathrm{mag}}$ strongly decreases with growing $T$ and therefore the
large estimated error $\Delta l_0=1000$~{\AA} significantly affects $l_h$ only at low $T$ (cf. Fig.~\ref{fig2}), playing no role in the following analysis.}
\bibitem{foot3}{We note that both $\rho_c$ and the electrical transport measured along the $a$-axis ($\rho_a$) are similarly correlated to $l_h$.}
\bibitem{foot4}Note that by this choice of derivatives it is not intended to imply any conclusions about activated transport in the material. However, with respect to other representations which similarly show the correlations between $\rho$ and $l_h$, the signatures of charge order are most clearly visible with this choice.
\end{thebibliography}
\end{document}